\documentclass[12pt,a4paper]{article}

\usepackage{graphicx}
\usepackage{amsmath}
\usepackage{amssymb}

\begin{document}

\title{\bf Thermo-charged capacitors and the Second Law of Thermodynamics}

\author{Germano D'Abramo\vspace{0.4cm}\\
{\small IASF - Istituto Nazionale di Astrofisica,}\\ 
{\small Via Fosso del Cavaliere 100,}\\
{\small 00133, Roma, Italy.}\\
{\small E--mail: {\tt Germano.Dabramo@iasf-roma.inaf.it}}\\
{\small Fax number: +39~06~49934188}}

\vspace{0.7cm}
\date{{\em Physics Letters A} {\bf 374} (2010) 1801}

\maketitle

\begin{abstract}

In this Letter we describe a vacuum spherical capacitor that generates a 
macroscopic voltage between its spheres harnessing the heat from a 
single thermal reservoir at room temperature. The basic idea is trivial 
and it makes use of two concentric spherical electrodes with different 
work functions. We provide a mathematical analysis of the underlying 
physical process and discuss its connections with the Second Law of 
Thermodynamics.\\

\noindent {\bf PACS (2010):} 79.40.+z, 67.30.ef, 65.40.gd, 84.32.Tt

\end{abstract}

\section{Introduction}

The Second Law of Thermodynamics, explicitly in the form of 
Kelvin-Planck postulate, puts a fundamental limit to the way in which 
usable work can be extracted from heat reservoirs, and to the maximum 
amount of work extractable (Carnot's principle). In particular, 
Kelvin-Planck postulate states that it is not possible to {\em 
cyclically} extract work from a {\em single} heat reservoir. Clausius 
postulate of the Second Law, which is notoriously equivalent to the 
Kelvin-Planck one, makes the impossibility more striking and 
understandable: heat can not spontaneously flow from sources at absolute 
temperature $T_1$ to sources at absolute temperature $T_2$ when $T_2\geq 
T_1$.

Although in the macroscopic physical world the Second Law seems 
authoritatively to make the difference between what is allowed and what 
is forbidden (to date, no experimental violation of the Second Law has 
been claimed), in the microscopic realm it seems to be continuously 
violated: let us take into account the Brownian motion or every 
fluctuation phenomena, for example.

About Brownian motion, Poincar\'e 
wrote~\cite{poin}:

\begin{quote}
``{\small [...] we see under our eyes now motion transformed into heat by 
friction, now heat changed inversely into motion, and that without loss 
since the movement lasts forever. This is contrary to the principle of 
Carnot.}''
\end{quote}

Almost every past attempt to understand and exploit such a microscopic 
violation relies on the approach of {\em fluctuations rectification}. 
Even the famous thought experiment of {\em Maxwell's Demon} is actually 
an idealized version of fluctuations rectification. The main 
difficulties which seem to afflict all these past approaches (sentient 
and non-sentient) are that every macroscopic/microscopic rectifier 
device seems either to suffer fluctuations itself, which neutralize 
every usable net effect, or its functioning seems to increase the total 
entropy of the system (at least of the same amount of the alleged 
reduction) mainly because of energy dissipation and/or entropy cost in 
the acquisition of information needed to run the device (for sentient 
devices).

For an interesting historical account of Second Law classical challenges 
(starting from Maxwell, and passing through Smoluchowski, Szilard, 
Brillouin, till Landauer) and their attempted resolutions, see Earman 
and Norton~\cite{eanor}.

Nonetheless, over the last 10-15 years an unparalleled number of 
challenges has been proposed against the status of the Second Law of 
Thermodynamics. During this period, more than 50 papers have appeared in 
the refereed scientific literature (see, for example, 
references~\cite{bib1,bib2,bib3, 
bib4,bib5,bib6,bib7,bib8,bib9,bib10,bib11,bib12,bib13,bib14,bib15,bib16, 
bib17,bib18,bib19}), together with a monograph entirely devoted to this 
subject~\cite{cs}. Moreover, during the same period of time two 
international conferences on the limit of the Second Law were also 
held~\cite{sh1,sh2}.

The general class of recent challenges~\cite{cs,bib16,bib19} spans 
plasma~\cite{bib15}, chemical~\cite{bib18}, gravitational~\cite{bib8} 
and solid-state physics~\cite{bib17,bib17b,bib19}. Currently, all these 
approaches appear immune to standard Second Law defenses (for a 
compendium of classical defenses, see~\cite{eanor} again) and several of 
them account laboratory corroboration of their underlying physical 
processes.

The present Letter aims to describe another approach to microscopic 
rectification that poses interesting theoretical questions along the 
aforementioned recent line of research: we are referring to an 
equivalent of the photo-electric effect with materials emitting 
electrons at room temperature. If we succeed in collecting all the 
electrons emitted by these materials in consequence of the absorption of 
black-body radiation from the uniformly heated environment, then we 
would be able to create a macroscopic voltage out of a single heat 
reservoir. Although this voltage appears to be {\em prima facie} an 
innocuous consequence of some well established physical laws, the whole 
process of thermo-charging presents quite interesting and paradoxical 
features which, analyzed with the tools of classical Thermodynamics, 
appear to violate the Second Law and seem to provide support to the 
recent results presented in the above cited literature.

The Letter is organized as follows: in the next section we describe our 
thermo-charged capacitor, which bears interesting similarities with some 
recent challenges presented in~\cite{bib15,bib16}, and provide a 
thorough mathematical analysis of its functioning. In the last section 
we discuss the paradoxical features of the physical process and show 
quantitatively why it seems to be at odds with the Second Law of 
Thermodynamics.

\section{Thermo-charged spherical capacitor}

\begin{figure}[h]

% LaTeX picture
\begin{center}
\begin{picture}(120,120)
\setlength{\unitlength}{0.5cm} \thicklines \put(4,6){\circle{1}}
\put(4,6){\circle{1.08}} \put(4,6){\circle{1.1}}
\put(4,6){\oval(4,4)}
\put(6.5,8.5){{\scriptsize Opaque case (with $\phi \gg 1\,$eV)}}
\put(0,8){{\scriptsize Vacuum}}
{\thinlines\put(5.5,7.65){\vector(1,1){0.8}}
\put(1.5,7.65){\vector(1,-1){1.2}}} \put(4,6.1){{\scriptsize $a$}}
\put(5,5.1){{\scriptsize $b$}}
{\thinlines\put(4,6){\line(1,0){0.5}}}
{\thinlines\put(4,6){\line(1,-1){1.6}}} \put(7,5){{\scriptsize Environment
$T\simeq 298\,$K}} {\thinlines
\put(3.5,6){\vector(-1,0){3}}} 
\put(-4,6){{\scriptsize Ag--O--Cs
sphere}} \put(-3.4,5){{\scriptsize $\phi\lesssim 0.7\,$eV}} 
{\thinlines \put(3,7.3){\vector(1,-1){0.8}}} {\thinlines
\put(3.8,6.5){\vector(1,1){0.8}}} \put(2.7,7.4){\scriptsize $h\nu$}
\put(4.5,7.4){\scriptsize $e^-$}
\end{picture}
\vspace{-1.5cm}
\end{center}
%\input{fig1.tex}
% LaTex picture
\caption{Scheme of the thermo-charged spherical capacitor with inner electrode
made of Ag--O--Cs (with $\phi_{in.}\lesssim 0.7\,$eV).}
\label{fig1}
\end{figure}
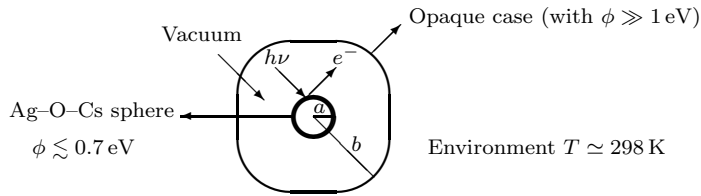

In Fig.~\ref{fig1} a sketched section of our vacuum spherical capacitor 
is shown. The outer sphere has radius $b$ and it is made of (conductive) 
material with high work function ($\phi_{ext.} \gg 1\,$eV). High work 
function means relatively low thermionic emission. The inner sphere, 
instead, has radius $a$ and it is made of a thermionic (conductive) 
material with relatively low work function ($\phi_{in.}\lesssim 1\,$eV). 
It should be clear that in such a case the two thermionic fluxes are 
different, the latter being greater than the former and the former 
negligible. Actually, this is true only at the beginning of the 
thermo-charging process; as a matter of fact, to reach our goal it 
suffices to concentrate on the initial phase of the process.

Let us describe how our device works. All the electrons emitted by the 
inner sphere, due to thermionic emission at room temperature, are
collected by the outer (very low emitting) sphere creating a macroscopic 
difference of potential $V$. Such a process lasts until $V$ is too high to 
be overcome by the kinetic energy $K_e$ of the main fraction of emitted 
electrons (namely, when $K_e < eV$, where $e$ is the charge of 
electron).

The work function of the outer sphere being greater than that of the 
inner one, $\phi_{ext.}\gg\phi_{in}$, the reverse process, namely the 
electronic flux from the outer sphere to the inner one, may be neglected 
without sensibly modifying our results. As said before, this is true 
only at the beginning of the charging process. At equilibrium, both 
fluxes are equal to that of the outer sphere and they balance each other 
exactly. Anyhow, our analysis concerns exclusively the initial phase of 
the process and the validity of our results remains unaffected by the 
establishment of the final equilibrium.

It could be interesting to provide an estimate of the value of $V$ 
obtainable and an estimate of the time needed to reach such a value, given 
the physical characteristics of the capacitor and the quantum efficiency 
curve $\eta(\nu)$ of thermionic material of the inner sphere.

The capacitor is placed in a heat bath at room temperature (environment) 
and it is subject to black-body radiation. Both spheres, at thermal 
equilibrium, emit and absorb an equal amount of radiation (Kirchhoff's 
law of thermal radiation), thus the amount of radiation absorbed by the 
inner sphere is the same as that emitted by that sphere according to the 
black-body radiation formula (Planck equation). Given the room 
temperature $T$, Planck equation provides us with the number 
distribution of photons absorbed as a function of their frequency.

According to the law of thermionic emission, the kinetic energy $K_e$ of 
the emitted electron is given by the following equation,

\begin{equation}
K_e=h\nu-\phi,
\label{eq1}
\end{equation}
where $h\nu$ is the energy of the photon with frequency $\nu$ ($h$ is 
the Planck constant) and $\phi$ is the work function of the material. 
Thus, only the tail of the Planck distribution of the absorbed photons, 
with frequency $\nu>\nu_0=\phi/h$, can contribute to the thermionic 
emission.

The voltage $V$ reachable with frequency $\nu_1$ is given by the 
following formula,

\begin{equation}
eV =h\nu_1-\phi,
\label{eq2}
\end{equation}
where $eV$ is the inter-sphere potential energy, and thus,

\begin{equation}
\nu_1=\frac{eV +\phi}{h}.
\label{eq3}
\end{equation}

The total number of photons per unit time $F_{p}$ with energy greater 
than or equal to $h\nu_1$, emitted and absorbed in thermal equilibrium 
by the inner sphere, is given by the Planck equation,

\begin{equation}
F_{p}=\frac{2\pi S}{c^2}\int_{\nu_1}^\infty
\frac{\nu^2 d\nu}{e^{\frac{h\nu}{kT}}-1},
\label{eq4}
\end{equation}
where $S$ is the inner sphere surface area, $c$ is the speed of light, 
$k$ is the Boltzmann constant and $T$ the room temperature.

If $\eta(\nu)$ is the quantum efficiency (or quantum yield) curve of 
thermionic material of the inner sphere, then the number of electrons 
per unit time $F_e$ thermionically emitted by the inner sphere towards 
the outer sphere with kinetic energy greater than or equal to 
$h\nu_1-\phi$, is given by

\begin{equation}
F_{e}=\frac{2\pi\cdot 4\pi a^2}{c^2}\int_{\nu_1}^\infty
\frac{\eta(\nu)\nu^2 d\nu}{e^{\frac{h\nu}{kT}}-1},
\label{eq5}
\end{equation}
where $4\pi a^2$ is the surface area of the inner sphere.

For a vacuum spherical capacitor, the voltage $V$ between the spheres 
and the charge $Q$ on each sphere are linked by the following equation,

\begin{equation}
V=\frac{Q}{4\pi \epsilon_0}\frac{b-a}{ab}.
\label{eq6}
\end{equation}

Now, we derive the differential equation which governs the process of 
thermo-charging. In the interval of time $dt$ the charge collected by 
the outer sphere is given by

\begin{equation}
dQ=eF_e dt=\frac{2\pi e\cdot 4\pi a^2}{c^2}\Biggl(\int_{\frac{eV(t) +
\phi}{h}}^\infty \frac{\eta(\nu)\nu^2
d\nu}{e^{\frac{h\nu}{kT}}-1}\Biggr)dt, 
\label{eq7}
\end{equation}
where we make use of eq.~(\ref{eq3}) for $\nu_1$ and $V(t)$ is the 
voltage at time $t$. Thus, through the differential form of 
eq.~(\ref{eq6}), we have

\begin{equation}
dV(t) = \frac{2\pi e}{\epsilon_0
c^2}\frac{a(b-a)}{b} \Biggl(\int_{\frac{eV(t) + \phi}{h}}^\infty
\frac{\eta(\nu)\nu^2 d\nu}{e^{\frac{h\nu}{kT}}-1}\Biggr)dt,
\label{eq8}
\end{equation}
or

\begin{equation}
\frac{dV(t)}{dt}= \frac{2\pi e}{\epsilon_0
c^2}\frac{a(b-a)}{b} \int_{\frac{eV(t) + \phi}{h}}^\infty
\frac{\eta(\nu)\nu^2 d\nu}{e^{\frac{h\nu}{kT}}-1}.
\label{eq9}
\end{equation}

Since our aim is to maximize the production of voltage $V$, we have to 
choose $a$ and $b$ such that they maximize the geometrical factor 
$a(b-a)/b$. It is not difficult to see that the maximum is reached when 
$a=b/2$. So we have

\begin{equation}
\frac{dV(t)}{dt}= \frac{\pi e b}{2\epsilon_0
c^2}\int_{\frac{eV(t) + \phi}{h}}^\infty \frac{\eta(\nu)\nu^2
d\nu}{e^{\frac{h\nu}{kT}}-1}.
\label{eq10}
\end{equation}

Unfortunately, provided that an analytical approximation of a real 
quantum efficiency curve $\eta(\nu)$ exists, the previous differential 
equation appears to have no general, simple analytical solution.

However, a close look at the Planckian integral of eq.~(\ref{eq10}) 
suggests to us the asymptotic behavior of $V(t)$. Even if we do not know 
{\em a priori} how $\eta(\nu)$ is, we know it to be a bounded function 
of frequency, with values between $0$ and $1$; usually, the higher is 
$\nu$, the closer to $1$ is $\eta(\nu)$. Thus, independently of 
$\eta(\nu)$, a slight increase of $V(t)$ makes the value of the 
Planckian integral to be smaller and smaller very fast. Heuristically, 
this suggests that $V(t)$ should tend quite rapidly to an `asymptotic' 
value (since $\frac{dV}{dt}$ tends to $0$).

In the rest of this section we provide a numerical solution of the above 
differential equation for the practical case of inner sphere made of 
Ag--O--Cs ~\cite{somm1, somm2, bates} (see Fig.~\ref{fig1}). To do that 
we need to adopt an approximation, however: the approximation consists 
in the adoption of a constant value for $\eta$, a sort of suitable mean 
value $\overline{\eta}$.

The differential equation~(\ref{eq10}) thus becomes

\begin{equation}
\frac{dV(t)}{dt}= \frac{\pi e b\overline{\eta}}{2\epsilon_0
c^2}\int_{\frac{eV(t) + \phi}{h}}^\infty \frac{\nu^2
d\nu}{e^{\frac{h\nu}{kT}}-1}.
\label{eq11}
\end{equation}

A straightforward variable substitution in the integral of 
eq.~(\ref{eq11}) allows to write it in its final, simplified form,

\begin{equation}
\frac{dV(t)}{dt}= \frac{\pi e b\overline{\eta}}{2\epsilon_0
c^2}\biggl(\frac{kT}{h}\biggr)^3\int_{\frac{eV(t) +
\phi}{kT}}^\infty \frac{x^2 dx}{e^{x}-1}.
\label{eq12}
\end{equation}

\begin{figure}[h]
\centerline{\includegraphics[width=7cm]{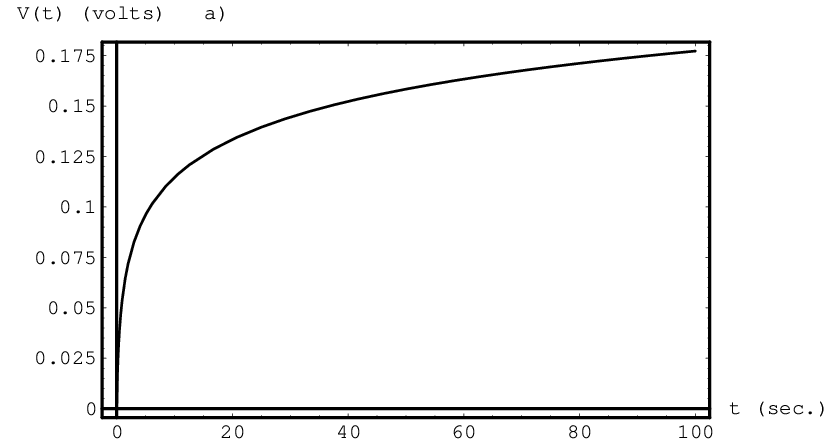}
  \includegraphics[width=7cm]{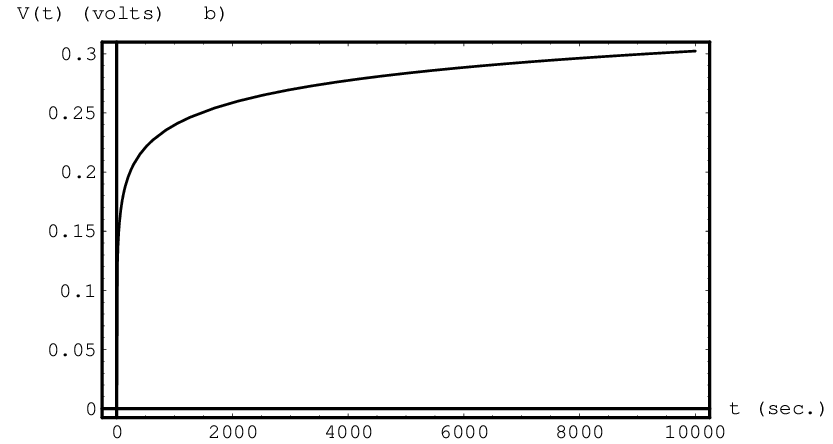}}
\caption{Thermo-charging process of the spherical capacitor described in 
the text ($\phi=0.7\,$eV, $b=0.20\,$m, $T=298\,$K, and 
$\overline{\eta}=10^{-5}$). These two plots show with different ranges 
in time-scale the behavior of $V(t)$. Plot (a) shows how only after 60 
seconds the voltage of the capacitor becomes more than $0.15\,$volts. 
Instead, plot (b) tells us that the voltage of the capacitor requires some
hours to approach $0.3\,$volts.}
\label{fig2}
\end{figure}

Here we provide an exemplificative numerical solution of 
eq.~(\ref{eq12}), adopting the following nominal values for $\phi$, $b$, 
$T$ and $\overline{\eta}$: $\phi=0.7\,$eV, $b=0.20\,$m, $T=298\,$K, and 
$\overline{\eta}=10^{-5}$. In order to make a conservative choice for 
the value of $\overline{\eta}$ we note that only black-body radiation 
with frequency greater than $\nu_0=\phi/h$ can contribute to thermionic 
emission. This means that for the Ag--O--Cs photo-cathode only radiation 
with wavelength smaller than $\lambda_0=hc/\phi\simeq 1700\,$nm 
contributes to the emission. According to Fig.~1 in Bates~\cite{bates}, 
the quantum yield of Ag--O--Cs for wavelengths smaller than $\lambda_0$ 
(and thus, for frequency greater than $\nu_0$) is always greater than 
$10^{-5}$.  Anyway, a laboratory realization of the capacitor, together 
with the experimental measurement of $V(t)$, should provide us with a 
realistic estimate of $\overline{\eta}$ for the Ag--O--Cs photo-cathode.

\begin{figure}[t]
\centerline{\includegraphics[width=7cm]{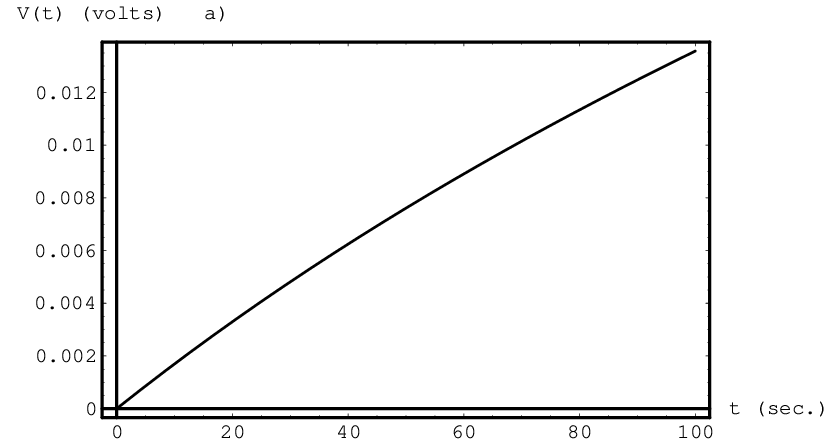}
  \includegraphics[width=7cm]{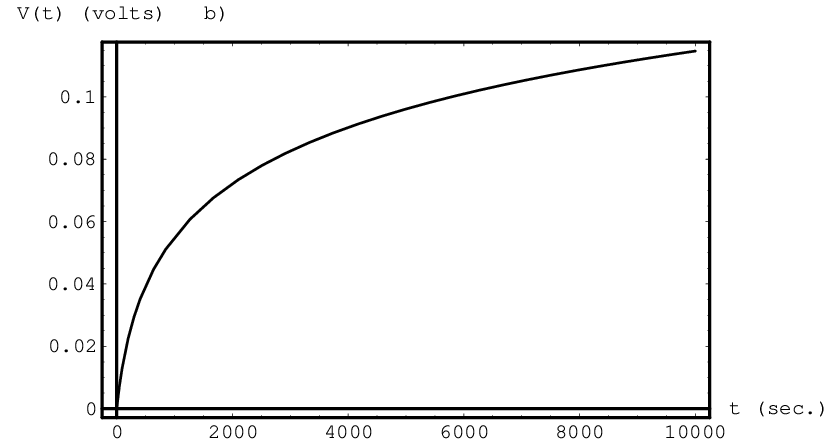}}
\caption{Thermo-charging process of the spherical capacitor described in 
the text with $\phi=0.7\,$eV, $b=0.20\,$m, $T=298\,$K, and 
$\overline{\eta}=10^{-8}$. These two plots show with different ranges in 
time-scale the behavior of $V(t)$. Plot (a) shows how only after 60 
seconds the voltage of the capacitor is near to $0.01\,$volts. Instead, 
plot (b) tells us that the voltage of the capacitor requires some
hours to become equal to $0.1\,$volts.}
\label{fig3}
\end{figure}

In Fig.~\ref{fig2} the numerical solution of the above test is shown. 
In plot (a) we could easily see how only after 60 seconds the voltage of 
the capacitor exceeds the value of $0.15\,$volts. Indeed, this is a 
macroscopic voltage. Instead, plot (b) tells us that the voltage of the 
capacitor requires some hours to approach $0.3\,$volts. Even in the more 
pessimistic scenario where $\overline{\eta}=10^{-8}$ we see that a 
macroscopic voltage should arise quite rapidly between the plates, see 
Fig.~\ref{fig3}.

\section{Discussion and conclusions}

At first sight, the charging process is a quite straightforward physical 
mechanism and it appears almost unproblematic. However, one feature of 
the thermo-charging should catch our attention. During the charging of 
the device the inner thermionic sphere substantially absorbs heat from 
the environment and releases this energy to the thermionic electrons. 
Such electrons fly to the outer sphere and impinge on it with non-zero 
velocity (since a non-zero fraction of them gathers their kinetic energy 
from the high energetic tail of the Planck distribution of black-body 
radiation). When they impinge on the outer sphere, they release their 
kinetic energy substantially heating the outer sphere. Thus, we are 
facing a spontaneous process involving an isolated system at uniform 
temperature (capacitor + environment), in which a part of the system 
(the inner sphere of the capacitor) absorbs heat at temperature $T$ and 
transfers a fraction of this heat to the other part of the system (the 
outer sphere) at the same temperature. This seems to macroscopically 
violate the Second Law of Thermodynamics in the Clausius formulation. As 
a matter of fact, if $Q_i$ is the energy absorbed by the inner sphere 
from the environment, $U$ is the energy stored in the electric field 
between the spheres ($U=\frac{1}{2}CV^2$, where $C=\frac{4\pi\epsilon_0 
ab} {b-a}$ is the capacitance of the spherical capacitor), and $Q_f$ is 
the energy transferred through the flying electrons to the outer sphere 
as heat (according to the First Law of Thermodynamics $Q_f+U=Q_i$, thus 
$Q_i>Q_f$), then the Clausius entropy variation of the whole system, as 
rough estimate, amounts to:

\begin{equation}
\Delta S_{tot}\simeq-\frac{Q_i}{T}+ \frac{Q_f}{T}<0. 
\label{eq13}
\end{equation}

In order to make the above result more striking, let us consider the 
following analogue in Classical Thermodynamics/Mechanics: a boulder of 
mass $m$ rests at the bottom of a valley, below a hill of height $h$, 
all the system at constant temperature $T$. Suddenly, the boulder 
spontaneously absorbs an amount $Q_1$ of heat (energy) from the 
environment and spontaneously starts to climb the hill at decreasing 
velocity (since the initial kinetic energy is being transformed into 
gravitational potential energy). Near the top of the hill the boulder 
hits a sort of wall and then stops. The friction experienced during the 
hit against the wall lets the boulder release to the environment an 
amount $Q_2$ of heat, obviously smaller than $Q_1$. According to the 
First Law of Thermodynamics we have: $Q_1-Q_2=mgh$, where $mgh$ is the 
gravitational potential energy variation of the boulder from the valley 
to the top of the hill.

Now, the total Clausius entropy variation is:

\begin{equation}
\Delta S_{tot}=-\frac{Q_1}{T}+\frac{Q_2}{T}=-\frac{mgh}{T}<0.
\label{eq14}
\end{equation}

The behavior of the boulder-environment system is practically the same 
as that of our electrons-environment system, and it is undoubtedly 
puzzling from the point of view of the Second Law of Thermodynamics.

Since the thermo-charging process is a spontaneous process, the 
variation of the Gibbs free energy of the system $\Delta G=\Delta H 
-T\Delta S$ should be negative. The enthalpy variation for constant 
pressure systems is given by $\Delta H =\Delta E + \Delta W$, where 
$\Delta E$ is the system internal energy variation and $\Delta W$ is the 
work produced. $\Delta H$ is equal to $0$ since, from the First Law of 
Thermodynamics, $\Delta H=\Delta Q =0$ (after all, we are dealing with 
an isolated system -- capacitor plus environment --, then the total 
$\Delta Q$ is equal to zero), and thus we have $\Delta G= -T\Delta S 
>0$. This is another way to see the paradoxical behavior of our system.

One possible objection could be that some other physical changes take 
place in the system during the process that may cancel out the apparent 
entropy decrease. Surely, during the charging process an electric field, 
and thus an electric potential, is created inside the capacitor where 
none existed before. Thus, one could attribute to the creation of the 
electric field an entropy increase greater than the entropy decrease due 
to the cooling and warming of the spheres.

A problem with this explanation could follow directly from the logic of 
the entropy variation analysis done for the ideal Carnot engine: the 
electric field represents a sort of work produced and stored in 
potential form by our capacitor, as could happen with a standard 
mechanical Carnot engine whose work is stored in a lifted weight. Hence, 
we are not able to see why for classical Carnot engine (and for 
Classical Thermodynamics) the work $W$ is not taken into account in the 
evaluation of the total entropy variation (unless it is transformed into 
heat, but it is not the case here), while for our thermo-charged 
capacitor the macroscopic stored energy (work) $U$ should be taken into 
account for the evaluation of the total $\Delta S$. In order to save the 
validity of the Second Law of Thermodynamics we need to find other 
sources of positive entropy during the charging process.

The above results do not seem to depend on the particular value of the 
work function of the outer sphere, provided that $\phi_{ext.} \gg 
1\,$eV. One may suppose that when the electrons fly into the outer 
sphere with higher work function, they diminish their potential energy 
and release a kinetic energy greater than that they had before impinging 
on the sphere, thus producing $Q_f>Q_i$ in eq.~(\ref{eq13}), and 
eventually giving $\Delta S>0$. But, what about the conservation of 
energy? As a matter of fact, if at some point of the charging process we 
try to neutralize both spheres putting them into contact at the same 
time with two distinct huge (infinite) chunks of neutral (conductive) 
materials, respectively of the same composition of the two spheres (in 
order to avoid problems with junction potentials) and at the same 
temperature $T$, then this process will release further energy through 
the Joule effect of the discharge (as a matter of fact, each sphere 
disperses its charge into the huge chunk of neutral material). This 
energy is equal to $U$, the energy of the thermo-generated electric 
field. The energy balance $\Delta E_{tot}$ of the entire process, 
thermo-charging and neutralization, thus gives: the total subtracted 
energy amounts to $Q_i$, the released energy amounts to $Q_f+U$, but 
according to the above attempted solution of the entropy paradox 
$Q_f>Q_i$, and thus $\Delta E_{tot}=Q_f+U-Q_i>U>0$, namely we would have 
a spontaneous energy production inside an isolated system. So, the above 
argument cannot be an acceptable explanation of the paradox.

This last incongruity sounds familiar, since it is not the first time 
that the Second Law is put against the First Law in scientific 
literature, see~\cite{bib17b}, p.~473.

Moreover, Sommerfeld in 1952~\cite{som}, trying to show how a {\em 
quasi-static} process does not always mean a {\em reversible} one, 
described the slow discharge of a charged capacitor through a very large 
resistance submersed in a heat reservoir: the discharge will take place 
by an arbitrarily small current, and negligible disturbance of 
electrostatic equilibrium, thus, such a process is {\em quasi-static}, 
but not {\em reversible}. In fact, the reverse process, he said, is not 
allowed by the Second Law~\cite{uff}.

To the author knowledge, our result represents an easily understandable 
and not so trivially refutable challenge to the Clausius formulation of 
the Second Law (see eq.~(\ref{eq13})). In its current incarnation, 
however, it is not completely clear how this alleged challenge could be 
made exploitable, namely how our capacitor could be transformed into a 
device able to {\em cyclically} produce usable net work\footnote{E.g., 
as DC current, or as electro-mechanical work like 
in~\cite{bib17,bib17b,bib19}. We can not reasonably rely on the 
inter-sphere heat flux alone, since it is microscopic and hence it can 
hardly be exploited thermodynamically/mechanically.} out of a single 
heat bath (e.g. challenging the Kelvin-Planck formulation of the Second 
Law).

As a matter of fact, if our result will resist the deep scrutiny of the 
scientific community in the years to come, then we are reasonably 
confident that, for our approach, an efficient path from ``Clausius 
violation'' to the ``Kelvin-Planck violation'' of the Second Law can be 
found, the two formulations being notoriously equivalent. This last 
subject is currently under investigation by the author.

\section*{Acknowledgments}

This work was partially supported by an Italian Space Agency (ASI) 
grant. The author acknowledges the instructive and supportive attitude 
of the Editor and the two anonymous referees. The author wishes also to 
gratefully acknowledge stimulating and encouraging discussions with 
Prof.~Daniel~P.~Sheehan (University of San Diego) during the preparation 
of an early draft of the manuscript.

\end{document}